\documentclass[preprint,12pt]{elsarticle}

\usepackage{amsmath}
\usepackage{amssymb}
\usepackage{booktabs}
\usepackage{array}
\usepackage{graphicx}
\usepackage{xcolor}
\usepackage{hyperref}
\usepackage{url}

\graphicspath{{figures/}}
\biboptions{sort&compress}
\journal{Applied Energy}

\begin{document}

\begin{frontmatter}

\title{Mobility-safe adaptive reserve certification for EV--hydrogen-bus--building resilience hubs}

\author[mcgill]{Yifan Wang\corref{cor1}}
\ead{yifan.wang18@mail.mcgill.ca}
\cortext[cor1]{Corresponding author.}

\affiliation[mcgill]{
  organization={Department of Mechanical Engineering, McGill University},
  city={Montreal},
  state={QC},
  postcode={H3A 2T7},
  country={Canada}
}

\begin{abstract}
Zero-emission mobility depots are becoming attractive resilience assets because the same site can host electric-vehicle charging, hydrogen-bus operation, stationary conversion equipment, and nearby critical-building backup. The operating problem is not simply how much energy can be exported during an outage. Hydrogen exported to a building can strand buses, electric-vehicle availability is stochastic, and critical-building demand changes seasonally. This paper introduces a mobility-safe reserve certification framework for a coupled electric-vehicle, hydrogen-bus, and critical-building hub. The framework combines a physics-hybrid universal differential equation building-load twin, one-sided split conformal reserve calibration, adaptive conformal inference for seasonal drift, and a mobility-first scheduling rule that protects post-event bus service before assigning hydrogen to buildings. The evaluation uses 495,221 real electric-vehicle charging sessions across eight reported regions, AC Transit GTFS-derived hydrogen-bus service days, and EnergyPlus 25.2 critical-building simulations under real TMY3 weather. Across 66,816 held-out outage scenarios, a mobility-blind hydrogen-export policy served 39.2\% of critical-building demand but protected buses in 0\% of cases and produced a 426.7 kg mean bus-hydrogen shortfall. A nominal mean-resource promise delivered only 45.4\% of commitments. The certified mobility-first policy was the only tested policy that simultaneously achieved 100\% commitment delivery, 100\% bus protection, and zero mean bus-hydrogen shortfall, while serving 20.5\% of critical-building demand. Under a summer-to-winter load shift, adaptive conformal inference raised late-period empirical coverage from 0.687 to 0.831 and reached 0.891 overall coverage against a 0.90 target with lower mean reserve than static split conformal. Across 12 building/seed drift runs, the adaptive conformal policy kept low late-coverage variability and the lowest mean reserve among the tested methods. The results show that resilience value in shared zero-emission hubs depends on service-aware certification, not on raw export capacity alone.
\end{abstract}

\begin{keyword}
critical buildings \sep vehicle-to-building \sep hydrogen buses \sep energy resilience \sep adaptive conformal inference \sep scientific machine learning
\end{keyword}

\end{frontmatter}

\section{Introduction}
\label{sec:introduction}

Power outages turn transportation energy infrastructure into a public-service allocation problem. An electric or hydrogen depot may appear to contain spare energy, but that energy is already attached to different obligations: buses must finish service, electric vehicles may or may not be connected when the event occurs, and critical buildings must sustain life-safety functions. Treating the depot as a generic battery can therefore create a hidden failure. A policy that maximizes building export can consume the hydrogen reserve needed for bus continuity, while a policy that commits only mean available resources can fail exactly on adverse days.

Recent work has shown that mobile energy resources can strengthen building and community resilience during outages. Electric vehicles and shared autonomous electric vehicles have been studied as flexible backup resources for critical buildings and mobility services \cite{tian2021resilience,liu2024critical,manzolli2026balancing}. Electric school buses have also been evaluated as backup sources for educational-function continuation \cite{liu2025schoolbusbackup}. In parallel, electric-bus depot scheduling, electric-vehicle aggregators, hybrid charging/refueling stations, and hydrogen-vehicle energy hubs have advanced rapidly \cite{jia2025multidepot,zhang2023decentralized,tahir2024hybrid,kazemi2025hydrogen,haddad2025integrated}. These streams make the same future plausible: a zero-emission depot will be asked to support both mobility and buildings during outages.

The unresolved operational question is how such a depot should make day-ahead reserve commitments when forecast errors have asymmetric service consequences. In a conventional energy-hub formulation, an optimistic hydrogen export may be counted as useful building service. In a resilience formulation, that same export is unacceptable if it strands hydrogen buses after the event. Static reserve margins also become fragile when the building-load distribution moves from an easier season into a harder one. Existing vehicle-to-building, bus-depot, hydrogen-hub, and building-backup studies generally analyze one or two parts of this system, but not a single mobility-first decision that certifies bus continuity and critical-building backup under non-stationary forecast error.

This paper addresses that gap with a mobility-safe adaptive reserve certification framework for an electric-vehicle (EV), hydrogen-bus, and critical-building hub. The framework has four linked components. First, a physics-hybrid universal differential equation (UDE) models critical-building demand as a thermal-inertia process with a neural forcing term \cite{rackauckas2020ude,gokhale2022pinn,raissi2019pinn}. Second, one-sided split conformal prediction converts point forecasts into service-specific upper reserve bounds under exchangeable calibration data \cite{shafer2008tutorial,lei2018distribution,borrotti2024conformal}. Third, adaptive conformal inference (ACI) updates the effective miscoverage level online when seasonal residuals drift \cite{gibbs2021adaptive,suresh2025mamba}. Fourth, a mobility-first certification rule protects post-event hydrogen-bus demand and station reserve before hydrogen is converted to building backup.

The central finding is that the most valuable resilience policy is not the one with the largest raw building export. Using real EV charging sessions, AC Transit GTFS-derived hydrogen-bus service requirements, and EnergyPlus critical-building loads, we show that mobility-blind hydrogen export gives an apparently high building-service fraction but completely fails the bus-continuity requirement. In contrast, the certified mobility-first policy sacrifices unsafe export, delivers every commitment in the held-out scenarios, and protects bus service in every tested case. Under seasonal building-load drift, the adaptive conformal layer restores coverage toward the target and reduces reserve variability across buildings and random seeds. These results reposition shared zero-emission depots as service-priority resilience systems, where statistically certified commitments matter more than nominal capacity.

\section{Related work and research gap}
\label{sec:related}

\subsection{Mobile energy resources for building resilience}

Vehicle-to-building and vehicle-to-grid studies have established that connected vehicles can provide meaningful backup energy during outages \cite{tian2021resilience,liu2024critical,manzolli2026balancing}. Recent work extends this idea to fleets, schools, and shared mobility systems \cite{liu2025schoolbusbackup}. The common decision is how to allocate available vehicle energy to buildings while preserving mobility utility. Our setting differs in two ways. First, the mobility service includes hydrogen buses, whose energy reserve is not interchangeable with EV connection availability. Second, the backup commitment is certified from data before the realized outage day, so feasibility is evaluated on held-out scenarios rather than assumed from nominal capacity.

\subsection{Electric-bus, hydrogen-vehicle, and multi-energy hub operation}

Depot-scale electrified transport has been studied through electric-bus scheduling and charging coordination, multi-energy hub optimization, and hybrid charging/refueling station design \cite{jia2025multidepot,zhang2023decentralized,tahir2024hybrid,kazemi2025hydrogen,haddad2025integrated}. These papers clarify infrastructure coupling and resource limitations, but their main objective is usually cost, scheduling feasibility, station design, or normal-operation coordination. In outage resilience, the same resource coupling becomes a service-priority problem: hydrogen must first cover bus service, and only certified surplus can support buildings. That service ordering is the main departure from mobility-blind energy-export formulations.

\subsection{Building-load forecasting and uncertainty certification}

Physics-informed and scientific-machine-learning models have improved building thermal modeling by combining structure with data-driven flexibility \cite{gokhale2022pinn,raissi2019pinn,rackauckas2020ude}. Conformal prediction provides distribution-free prediction sets under exchangeability and has been used to quantify uncertainty in load and building-performance settings \cite{shafer2008tutorial,lei2018distribution,borrotti2024conformal}. However, outage reserve decisions are often exposed to seasonal distribution shift, where a static calibration margin can be too small in later high-demand periods. ACI was designed for online adjustment under distribution shift \cite{gibbs2021adaptive}, and recent energy forecasting work has begun to integrate adaptive conformal methods \cite{suresh2025mamba}. The missing link is to connect adaptive coverage control directly to service-aware EV--hydrogen-bus--building reserve certification.

\section{Problem formulation}
\label{sec:problem}

We consider a hub that must commit critical-building backup before the realized outage day is known. Let $t$ index days, $r$ EV data regions, $b$ critical-building types, and $d \in \{6,12,24\}$ outage duration in hours. The uncertain quantities are EV connection-derived availability $E^{\mathrm{EV}}_{r,t,d}$, hydrogen tank state at the event $H^{\mathrm{tank}}_{t}$, post-event bus hydrogen requirement $H^{\mathrm{bus}}_{t}$, and critical-building energy demand $Y_{b,t,d}$. A policy $\pi$ announces a building-backup commitment $C^{\pi}_{r,t,b,d}$ and then faces the realized resources.

The hub is evaluated by three service metrics. The critical-building service fraction is
\begin{equation}
S^{\pi}_{r,t,b,d} =
\frac{\min\{C^{\pi}_{r,t,b,d},Y_{b,t,d}\}}{Y_{b,t,d}},
\label{eq:service}
\end{equation}
with $Y_{b,t,d}>0$. Commitment delivery is the event that the realized available energy is at least the promised commitment:
\begin{equation}
D^{\pi}_{r,t,b,d} =
\mathbb{I}\{E^{\mathrm{EV}}_{r,t,d}+E^{\mathrm{H2}\rightarrow \mathrm{B}}_{t,d}+E^{\mathrm{ride}}_{b,d}
\ge C^{\pi}_{r,t,b,d}\}.
\label{eq:delivery}
\end{equation}
Bus protection is the event that enough hydrogen remains after building export to satisfy post-event bus service and station reserve:
\begin{equation}
P^{\pi}_{t,d} =
\mathbb{I}\{H^{\mathrm{tank}}_{t} - H^{\mathrm{used}}_{t,d}
\ge H^{\mathrm{bus}}_{t}+H^{\mathrm{floor}}\}.
\label{eq:busprotect}
\end{equation}
The objective is not to maximize Eq.~\eqref{eq:service} alone. A resilience-feasible policy must keep $D^{\pi}$ and $P^{\pi}$ high while serving as much critical-building demand as the certified resources allow.

\section{Mobility-safe Sci-ML reserve certification}
\label{sec:method}

\begin{figure}[t]
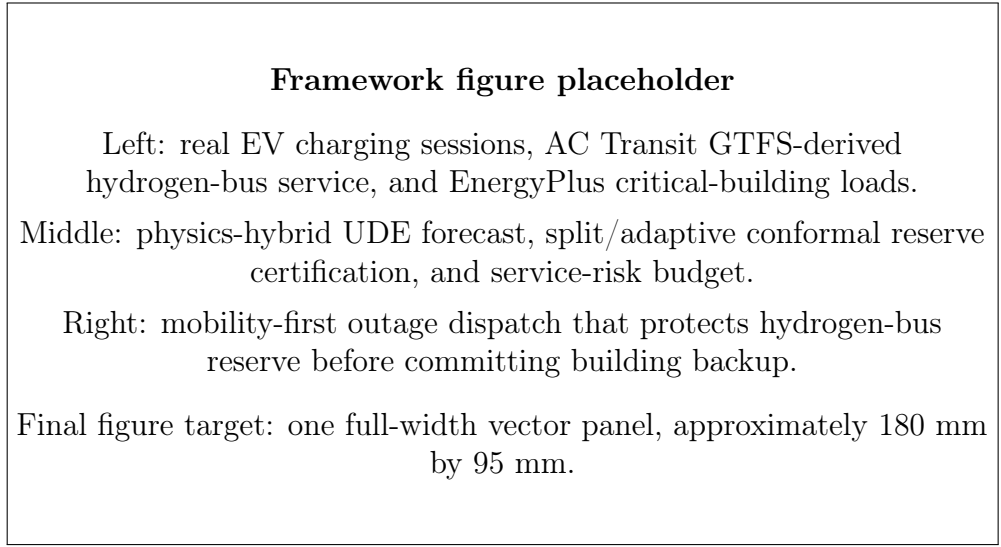

\centering
\fbox{\begin{minipage}[c][0.36\textheight][c]{0.94\linewidth}
\centering
\textbf{Framework figure placeholder}\\[0.8em]
Left: real EV charging sessions, AC Transit GTFS-derived hydrogen-bus service, and EnergyPlus critical-building loads.\\[0.4em]
Middle: physics-hybrid UDE forecast, split/adaptive conformal reserve certification, and service-risk budget.\\[0.4em]
Right: mobility-first outage dispatch that protects hydrogen-bus reserve before committing building backup.\\[0.8em]
Final figure target: one full-width vector panel, approximately 180 mm by 95 mm.
\end{minipage}}
\caption{Planned framework schematic for the EV--hydrogen-bus--building resilience hub. The final artwork will replace this placeholder with a graphical flow from data streams to certified outage dispatch.}
\label{fig:framework}
\end{figure}

Figure~\ref{fig:framework} gives the intended paper-level schematic. The implemented framework links three evidence streams to four computational modules: physical load forecasting, conformal reserve calibration, adaptive drift response, and mobility-first dispatch certification.

\subsection{Physics-hybrid building-load twin}

For each building type $b$, the daily critical load trajectory is represented as a first-order relaxation process,
\begin{equation}
\frac{dL_{b,t}(\tau)}{d\tau}
= \kappa_b\left[f_{\theta_b}(\tau,m_t,z_{b,t})-L_{b,t}(\tau)\right],
\qquad
L_{b,t}(0)=L^{0}_{b,t},
\label{eq:ude}
\end{equation}
where $\tau$ is hour within the day, $m_t$ is month, $z_{b,t}$ contains contextual inputs, $\kappa_b>0$ is a learned relaxation rate, and $f_{\theta_b}$ is a neural forcing term. The model is a UDE because known thermal inertia is retained while unmodeled forcing is learned from data. The outage energy forecast is
\begin{equation}
\widehat{Y}_{b,t,d}=\sum_{\tau \in \mathcal{W}_{t,d}}\widehat{L}_{b,t}(\tau)\Delta \tau,
\label{eq:energyforecast}
\end{equation}
where $\mathcal{W}_{t,d}$ is the outage window and $\Delta \tau=1$ h in the implementation.

The critical-building profiles are derived from EnergyPlus hourly end uses. For building $b$, the critical load is constructed as
\begin{equation}
L^{\mathrm{crit}}_{b,t} =
L^{\mathrm{cool}}_{b,t}
 + L^{\mathrm{fan}}_{b,t}
 + L^{\mathrm{pump}}_{b,t}
 + L^{\mathrm{heat}}_{b,t}
 + L^{\mathrm{water}}_{b,t}
 + \rho_b\left(L^{\mathrm{light}}_{b,t}+L^{\mathrm{equip}}_{b,t}\right),
\label{eq:critical}
\end{equation}
with $\rho_b=0.60$ for the critical hospital, $0.40$ for the cooling center, and $0.35$ for the warming center and mass-care shelter. The coefficient $\rho_b$ represents the life-safety fraction of lighting and equipment load retained during outage operation.

\subsection{One-sided conformal reserve bounds}

Let $\mathcal{I}_{\mathrm{cal}}$ be a calibration set with true demand $y_i$ and point forecast $\widehat{y}_i$. The one-sided nonconformity score is
\begin{equation}
s_i = y_i-\widehat{y}_i,\qquad i\in\mathcal{I}_{\mathrm{cal}}.
\label{eq:score}
\end{equation}
For a target miscoverage $\alpha$, the finite-sample corrected quantile level is
\begin{equation}
\ell_{\alpha} = \min\left\{1,\frac{\lceil (n_{\mathrm{cal}}+1)(1-\alpha)\rceil}{n_{\mathrm{cal}}}\right\},
\label{eq:level}
\end{equation}
and the upper reserve bound is
\begin{equation}
U_{t}(\alpha)=\widehat{y}_{t}+Q_{\ell_{\alpha}}\left(\{s_i:i\in\mathcal{I}_{\mathrm{cal}}\}\right).
\label{eq:split}
\end{equation}
This construction is applied to building backup energy and bus hydrogen demand. The total service-risk budget $\alpha$ is split across services with relative laxity weights $w_j$:
\begin{equation}
\alpha_j = \alpha\frac{w_j}{\sum_k w_k},
\label{eq:riskbudget}
\end{equation}
so the mobility service receives the tighter allocation by assigning it the smaller weight. In the reported conformal snapshot, the implementation uses weights $w_{\mathrm{mobility}}=0.5$ and $w_{\mathrm{building}}=1.5$.

\subsection{Adaptive conformal inference under seasonal drift}

Static split conformal uses a fixed calibration distribution. To track seasonal residual changes, ACI maintains a time-varying effective miscoverage level $a_t$. With past residuals $\mathcal{S}_{t-1}=\{y_i-\widehat{y}_i:i<t\}$, the online bound is
\begin{equation}
U_t^{\mathrm{ACI}}=\widehat{y}_t+Q_{1-a_t}(\mathcal{S}_{t-1}).
\label{eq:acibound}
\end{equation}
After the realization is observed, the update is
\begin{equation}
a_{t+1}=\Pi_{[10^{-3},0.5]}\left(a_t+\gamma\left[\alpha-\mathbb{I}\{y_t>U_t^{\mathrm{ACI}}\}\right]\right),
\label{eq:aciupdate}
\end{equation}
where $\Pi$ clips the value to the feasible interval. A violation decreases $a_t$, raises the next quantile level, and widens the reserve. Repeated safe days relax the reserve. The reported drift experiment uses $\alpha=0.10$, $\gamma=0.20$, and a 60-day warm-up for the illustrative summer-to-winter stream.

\subsection{Mobility-first hydrogen and building dispatch}

Hydrogen can support buildings only after bus service is protected. Let $e_{\mathrm{H2}}=33.33$ kWh kg$^{-1}$ be the hydrogen lower heating value, $\eta_{\mathrm{fc}}=0.50$ the stationary fuel-cell efficiency, $\eta_{\mathrm{exp}}=0.12$ the export fraction used in the hub model, $H^{\mathrm{floor}}=250$ kg the station floor, and $P_{\mathrm{fc}}=250$ kW the stationary fuel-cell power. The certified post-event bus reserve is
\begin{equation}
R^{\mathrm{bus}} =
U^{\mathrm{bus}}(\alpha_{\mathrm{mobility}})+H^{\mathrm{floor}},
\label{eq:busreserve}
\end{equation}
and the certified tank inventory is a lower quantile of calibration tank states,
\begin{equation}
\underline{H}^{\mathrm{tank}} = Q_{\epsilon}\left(\{H^{\mathrm{tank}}_i:i\in\mathcal{I}_{\mathrm{cal}}\}\right).
\label{eq:tanklower}
\end{equation}
The hydrogen mass available for building export is therefore
\begin{equation}
H^{\mathrm{exp}}=\left[\underline{H}^{\mathrm{tank}}-R^{\mathrm{bus}}\right]_{+},
\label{eq:hexport}
\end{equation}
and the corresponding building energy is
\begin{equation}
E^{\mathrm{H2}\rightarrow\mathrm{B}}_{d} =
\min\left\{H^{\mathrm{exp}}e_{\mathrm{H2}}\eta_{\mathrm{fc}}\eta_{\mathrm{exp}},\,
P_{\mathrm{fc}}d\right\}.
\label{eq:h2energy}
\end{equation}
Finally, the certified building commitment is
\begin{equation}
C^{\mathrm{cert}}_{r,t,b,d}
=
\min\left\{
U^{\mathrm{building}}_{b,t,d},
E^{\mathrm{EV,cert}}_{r,d}+E^{\mathrm{H2}\rightarrow\mathrm{B}}_{d}+E^{\mathrm{ride}}_{b,d}
\right\},
\label{eq:commitment}
\end{equation}
where $E^{\mathrm{ride}}_{b,d}$ is a conservative 1.5 h thermal ride-through term estimated from the seasonal mean critical load.

\section{Data, simulation, and experimental design}
\label{sec:data}

\subsection{Evidence streams}

The study combines three evidence streams. The EV stream contains 495,221 standardized charging sessions across eight reported regions, assembled from open EV charging datasets including Town of Cary and EV Load Open Data sources \cite{townofcarydata,evloadopendata}. The EV variable used in the experiment is connection-derived outage-window availability. It measures the energy that would be available if connected vehicles participate in a managed resilience program.

The hydrogen-bus stream is derived from the AC Transit GTFS data API \cite{actransitdata}. Daily bus-service distances are converted into hydrogen requirements using 0.115 kg km$^{-1}$. The depot model uses a 30-bus hydrogen fleet, 1500 kg station storage, 70\% initial station state of charge, a 42 kg h$^{-1}$ electrolyzer rate, and a 17:00--20:00 event window. The resulting table contains 56 service days with daily hydrogen demand, tank state at event time, post-event remaining bus demand, and hydrogen mass potentially available for building export.

The building stream is generated with EnergyPlus 25.2 using ASHRAE-90.1 prototype-style critical-building profiles and real TMY3 weather \cite{crawley2001energyplus,energyplus2026}. Four building roles are included: critical hospital, cooling center, warming center, and mass-care shelter. These profiles supply hourly critical loads for outage windows and daily trajectories for the UDE and conformal experiments.

\begin{figure}[t]
\centering
\includegraphics[width=\linewidth]{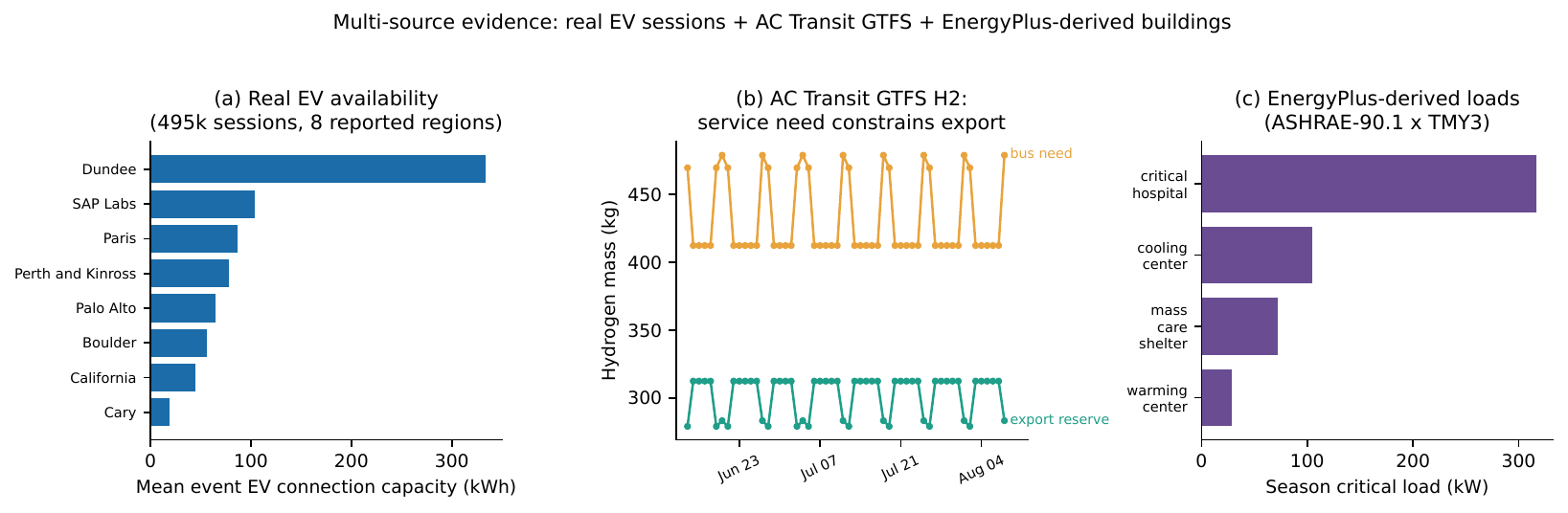}
\caption{Data and system evidence for the coupled hub. The figure summarizes EV connection-derived availability, hydrogen-bus reserve structure, and EnergyPlus-derived critical-building load profiles.}
\label{fig:data}
\end{figure}

\subsection{Baselines and metrics}

The scheduling experiment compares six operating points. The building-only baseline uses thermal ride-through alone. The EV-plus-building baseline adds certified EV availability but no hydrogen export. The mobility-blind hydrogen baseline exports hydrogen after only a station floor and ignores remaining bus service. The nominal tri-vector baseline commits from mean EV, hydrogen, and building ride-through resources. The certified mobility-first method uses quantile-certified EV availability, bus reserve, tank inventory, and building backup. The certified CVaR variant replaces quantiles with expected shortfall in the adverse tail.

The main metrics are mean critical-building service, fifth-percentile critical-building service, commitment delivery rate, bus protection rate, mean bus shortfall, and priority-weighted service. The conformal experiments additionally report empirical coverage and mean reserve. For the seasonal-drift experiment, ``early'' and ``late'' coverage split the test stream into two equal halves after calibration.

\section{Results}
\label{sec:results}

\subsection{Mobility-blind export creates a hidden bus-service failure}

Table~\ref{tab:scheduling} reports 66,816 held-out scheduling scenarios across eight EV regions and three outage durations. Mobility-blind hydrogen export achieved the highest mean critical-building service, 39.2\%, but protected buses in 0\% of cases and created a 426.7 kg mean bus-hydrogen shortfall. This is the central operational failure: raw building export can be high precisely because it consumes the reserve needed for bus continuity.

The nominal tri-vector policy also failed the resilience test. It served 27.6\% of critical-building demand on average but delivered only 45.4\% of its commitments. In contrast, certified mobility-first dispatch delivered 100\% of commitments, protected buses in 100\% of cases, and produced zero mean bus-hydrogen shortfall while serving 20.5\% of critical-building demand. The certified CVaR variant showed nearly identical feasibility and slightly lower service. The result identifies a practical operating frontier: unsafe export looks attractive in an energy-only metric, but service-aware certification is required for feasible resilience operation.

\begin{table}[t]
\centering
\caption{Held-out scheduling results over 66,816 scenarios. Service values are fractions of critical-building demand; delivery and bus protection are empirical rates.}
\label{tab:scheduling}
\small
\resizebox{\linewidth}{!}{%
\begin{tabular}{lccccc}
\toprule
Method & Cases & Critical service & Delivery & Bus protected & Bus shortfall (kg)\\
\midrule
Building ride-through & 11136 & 0.106 & 1.000 & 1.000 & 0.0\\
EV + building & 11136 & 0.109 & 1.000 & 1.000 & 0.0\\
Mobility-blind H2 export & 11136 & 0.392 & 1.000 & 0.000 & 426.7\\
Nominal tri-vector & 11136 & 0.276 & 0.454 & 1.000 & 0.0\\
Certified mobility-first & 11136 & 0.205 & 1.000 & 1.000 & 0.0\\
Certified CVaR & 11136 & 0.203 & 1.000 & 1.000 & 0.0\\
\bottomrule
\end{tabular}
}
\end{table}

\begin{figure}[t]
\centering
\includegraphics[width=\linewidth]{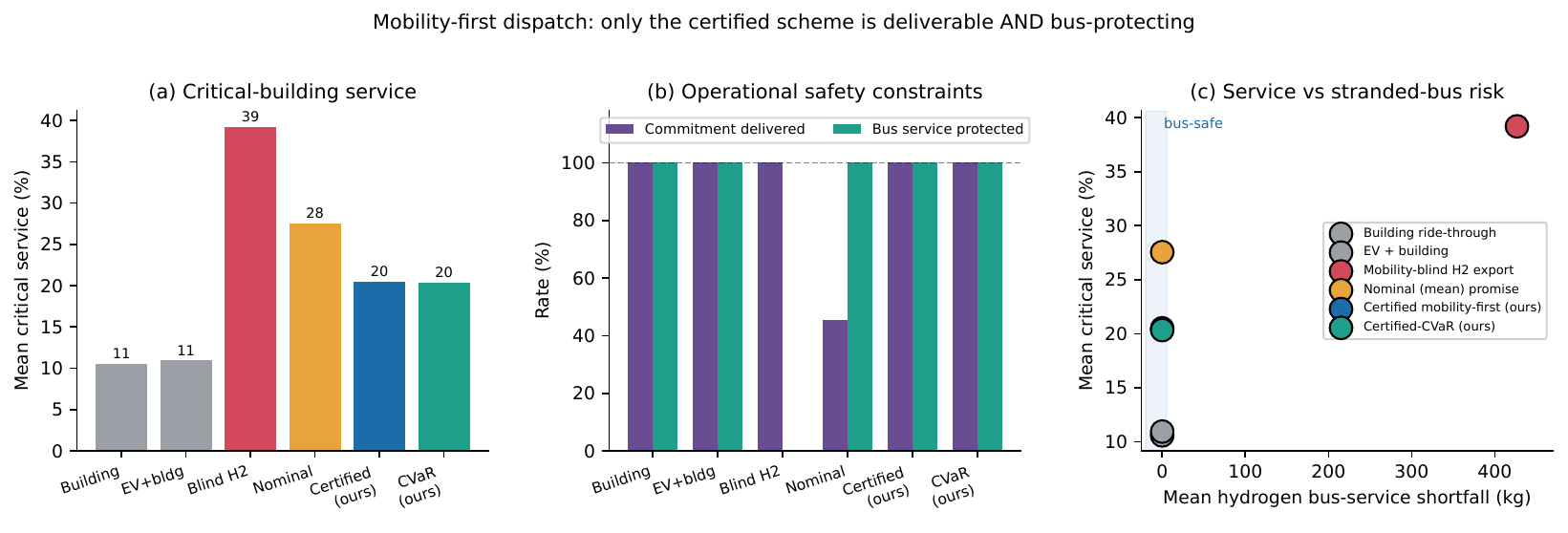}
\caption{Operational tradeoff between apparent building service and mobility-safe delivery. Mobility-blind hydrogen export gives high apparent building support but produces bus-service shortfall. Certified mobility-first dispatch lowers unsafe export and occupies the feasible, bus-protecting region.}
\label{fig:tradeoff}
\end{figure}

\subsection{The physics-hybrid twin is competitive and supports traceable reserve certification}

Table~\ref{tab:ude} compares the UDE, a black-box neural network, and persistence on held-out unseen-season EnergyPlus profiles. The UDE was competitive with the black-box model and consistently improved over persistence. For example, the UDE reduced RMSE from 56.6 to 25.5 kW for the cooling center and from 133.6 to 63.2 kW for the critical hospital. The black-box model was slightly more accurate for the warming center, mass-care shelter, and hospital. Therefore, the UDE contribution is not a pure forecasting leaderboard claim. Its role is to keep thermal inertia explicit while providing forecasts that are accurate enough for downstream reserve certification.

\begin{table}[t]
\centering
\caption{Held-out clean-season forecast RMSE for critical-building hourly load (kW).}
\label{tab:ude}
\begin{tabular}{lccc}
\toprule
Building & UDE & Black-box NN & Persistence\\
\midrule
Cooling center & 25.5 & 25.8 & 56.6\\
Warming center & 24.0 & 21.1 & 30.4\\
Mass-care shelter & 57.2 & 56.8 & 92.2\\
Critical hospital & 63.2 & 61.8 & 133.6\\
\bottomrule
\end{tabular}
\end{table}

\subsection{Conformal reserves convert brittle point forecasts into service coverage}

The held-out conformal snapshot shows why point forecasts alone are not sufficient for outage commitments. With target coverage 0.90, the point forecast protected buses in 0.696 of cases and covered building backup in 0.391 of cases. A fixed empirical quantile margin reached 1.000 bus protection and 0.913 building coverage. The split conformal service-risk budget reached 1.000 bus protection and 0.978 building coverage, with mean reserves of 1235 kg for bus hydrogen and 10,878 kWh for building backup. This result supports the paper's use of conformal reserve certification as a forecast-to-commitment layer.

\begin{table}[t]
\centering
\caption{Held-out conformal reserve snapshot for bus hydrogen and critical-building backup.}
\label{tab:conformal}
\small
\resizebox{\linewidth}{!}{%
\begin{tabular}{lcccc}
\toprule
Method & Target & Bus protected & Building covered & Building reserve (kWh)\\
\midrule
Point forecast & 0.90 & 0.696 & 0.391 & 8841\\
Empirical quantile & 0.90 & 1.000 & 0.913 & 10043\\
Split conformal budget & 0.90 & 1.000 & 0.978 & 10878\\
\bottomrule
\end{tabular}
}
\end{table}

\subsection{Adaptive conformal reserves respond to summer-to-winter drift}

\begin{figure}[t]
\centering
\includegraphics[width=\linewidth]{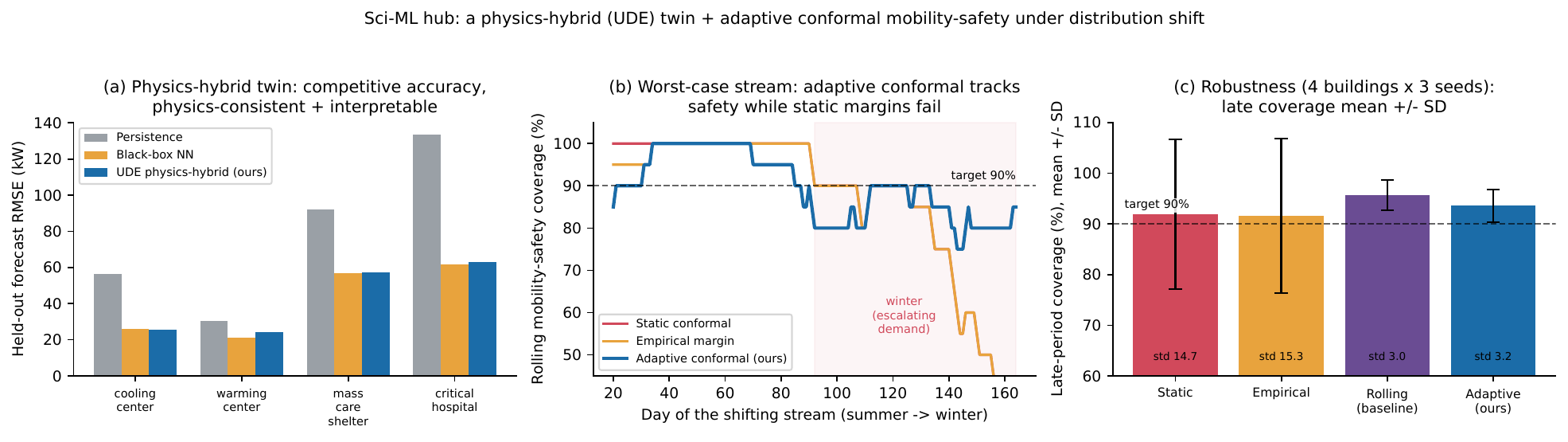}
\caption{Forecasting and reserve safety. Panel (a) compares the physics-hybrid UDE, black-box neural network, and persistence baselines. Panel (b) shows rolling coverage under the summer-to-winter drift stream. Panel (c) summarizes drift robustness across four buildings and three UDE seeds; error bars denote standard deviation across the 12 runs.}
\label{fig:sciml}
\end{figure}

Figure~\ref{fig:sciml} and Table~\ref{tab:shift} report the seasonal-drift stress test. The illustrative stream uses the warming-center profile rolled from an easier summer regime into a harder winter regime. Static split conformal achieved 1.000 early coverage but fell to 0.687 late coverage. A fixed empirical margin showed the same late-period failure. ACI raised late-period coverage to 0.831 and overall coverage to 0.891 against a 0.90 target. It also used lower mean reserve than static split conformal, 1296 versus 1319 kWh.

\begin{table}[t]
\centering
\caption{Summer-to-winter seasonal-drift stress test for daily critical-building backup.}
\label{tab:shift}
\small
\resizebox{\linewidth}{!}{%
\begin{tabular}{lcccc}
\toprule
Method & Overall coverage & Early coverage & Late coverage & Mean reserve (kWh)\\
\midrule
Static split conformal & 0.842 & 1.000 & 0.687 & 1319\\
Empirical fixed margin & 0.836 & 0.988 & 0.687 & 1292\\
Adaptive conformal & 0.891 & 0.951 & 0.831 & 1296\\
\bottomrule
\end{tabular}
}
\end{table}

\subsection{Robustness across buildings and seeds}

The drift experiment was repeated across all four EnergyPlus building profiles and three UDE seeds, with each stream ordered from the building's own low-demand month into its harder season. Table~\ref{tab:robust} includes a fair rolling empirical adaptive baseline. Static split conformal and fixed empirical margins had high late-coverage standard deviations, approximately 0.15, meaning their mean performance hid severe per-stream failures. Both adaptive methods reduced late-coverage variability to approximately 0.03. The rolling empirical adaptive baseline achieved the highest mean late coverage, 0.957, while ACI achieved 0.936 late coverage and the lowest mean reserve, 4622 kWh. The empirical advantage of ACI is therefore reserve efficiency and a principled feedback update, not a claim that it dominates every adaptive heuristic on raw coverage.

\begin{table}[t]
\centering
\caption{Robust seasonal-drift comparison over four buildings and three UDE seeds. Values are mean $\pm$ standard deviation across 12 runs.}
\label{tab:robust}
\small
\resizebox{\linewidth}{!}{%
\begin{tabular}{lccc}
\toprule
Method & Late coverage & Overall coverage & Mean reserve (kWh)\\
\midrule
Static split conformal & $0.919 \pm 0.147$ & $0.855 \pm 0.193$ & 5080\\
Empirical fixed margin & $0.916 \pm 0.153$ & $0.843 \pm 0.196$ & 4964\\
Rolling empirical adaptive & $0.957 \pm 0.030$ & $0.919 \pm 0.030$ & 4676\\
Adaptive conformal & $0.936 \pm 0.032$ & $0.914 \pm 0.030$ & 4622\\
\bottomrule
\end{tabular}
}
\end{table}

\section{Discussion}
\label{sec:discussion}

The experiments show that depot resilience must be judged by service feasibility rather than export volume. Mobility-blind hydrogen export is the clearest example. It gives the highest building-service fraction in Table~\ref{tab:scheduling}, but it does so by using hydrogen that the buses still need. From an energy-only perspective, that policy looks attractive. From a public-service perspective, it is infeasible. The certified mobility-first policy changes the decision order: reserve hydrogen for buses first, then convert only certified surplus to building backup.

The adaptive conformal results show a second reason why reserve certification is necessary. Static margins can look conservative in early low-demand periods yet under-cover in later high-demand periods. The warming-center stress test makes this failure visible, while the 12-run robustness table shows that the problem is not a single random seed. The adaptive methods keep much lower coverage variability than the static methods, and ACI obtains this reliability with the lowest mean reserve among the tested methods.

The UDE module contributes physical structure rather than universal forecast dominance. This distinction matters for an applied energy manuscript. A black-box predictor can be slightly more accurate on some building profiles, but the downstream safety layer is forecaster-agnostic: any point forecast can be wrapped in the conformal reserve. The UDE is useful because it makes the building-load twin interpretable and keeps thermal inertia explicit in a setting where reserve commitments must be explained to operators.

Several implementation choices should guide practical use. EV availability in this study is connection-derived and should be implemented with a managed participation program before being treated as dispatchable discharge. Hydrogen-bus demand is derived from one agency's GTFS service data and should be expanded to additional agencies as data access improves. EnergyPlus-derived building loads provide reproducible critical-load profiles under real weather; site deployment would replace them with meter-calibrated building models where available. These choices do not change the main operating lesson: shared zero-emission depots need service-priority reserve certification before they are used as outage backup assets.

\section{Conclusions}
\label{sec:conclusion}

This paper formulated and evaluated a mobility-safe reserve certification framework for a coupled EV, hydrogen-bus, and critical-building resilience hub. The framework combines a physics-hybrid building-load twin, split conformal reserve calibration, adaptive conformal drift response, and a mobility-first hydrogen allocation rule. Across 66,816 held-out scenarios, policies that ignored service priority either stranded buses or failed to deliver their commitments. The certified mobility-first policy was the only tested policy that protected buses, delivered every commitment, and provided nonzero critical-building backup. Under seasonal drift, ACI improved late-period coverage and reduced reserve variability while retaining the lowest mean reserve across the multi-building robustness runs. The result is a deployable planning principle for zero-emission depots: outage backup should be committed from certified service surplus, not from nominal energy capacity.

\section*{Data and code availability}

The manuscript is based on the local V3.2 package. The processed EV sessions, GTFS-derived hydrogen-bus reserve table, EnergyPlus-derived building load profiles, experiment scripts, result tables, and figure-generation scripts are organized under \texttt{data/}, \texttt{code/}, and \texttt{results/}. Public sources are cited in the bibliography. Final public release will remove local absolute paths and include scripts for regenerating the processed tables from public inputs where licensing permits.

\section*{Declaration of competing interest}

The author declares no known competing financial interests or personal relationships that could have appeared to influence the work reported in this paper.

\section*{Acknowledgements}

The author thanks the maintainers of the public EV charging, transit GTFS, EnergyPlus, and weather datasets used to build the pre-manuscript evidence package.

\bibliographystyle{elsarticle-num}
\bibliography{references}

@article{manzolli2026balancing,
  title = {Balancing energy resilience and mobility: a multi-objective strategy for deploying shared autonomous electric vehicles during power outages},
  author = {Manzolli, Jos{\'e} Augusto and Yu, Sibo and D'Apice, Danilo and Miranda-Moreno, Luis F.},
  journal = {npj Sustainable Mobility and Transport},
  year = {2026},
  doi = {10.1038/s44333-026-00081-9}
}

@article{jia2025multidepot,
  title = {Multi-depot battery electric bus scheduling and charging coordination under resource limitations},
  author = {Jia, Yan and An, Kai},
  journal = {Applied Energy},
  year = {2025},
  doi = {10.1016/j.apenergy.2025.126444}
}

@article{liu2025schoolbusbackup,
  title = {Resilience and environmental benefits of electric school buses as backup power for educational functions continuation during outages},
  author = {Liu, Hongyu and Vlachokostas, Alexandros and Kontou, Eleftheria},
  journal = {Building and Environment},
  year = {2025},
  doi = {10.1016/j.buildenv.2024.112329}
}

@article{kazemi2025hydrogen,
  title = {Hydrogen vehicle operational management in the context of green-hydrogen energy hub applications},
  author = {Kazemi, S. and Khalili, T. and Beydaghi, S. and Moeini-Aghtaie, M.},
  journal = {International Journal of Hydrogen Energy},
  year = {2025},
  doi = {10.1016/j.ijhydene.2025.152359}
}

@article{haddad2025integrated,
  title = {Integrated sizing and management of residential energy systems for electric and hydrogen vehicle charging},
  author = {Haddad, R. and Asadi, S. and Alemazkoor, N.},
  journal = {Journal of Building Engineering},
  year = {2025},
  doi = {10.1016/j.jobe.2025.114675}
}

@article{suresh2025mamba,
  title = {Mamba based adaptive conformal inference for probabilistic short-term load forecasting},
  author = {Suresh, V. and Swain, A. K. and Revathi, S. and Guerrero, J. M.},
  journal = {Knowledge-Based Systems},
  year = {2025},
  doi = {10.1016/j.knosys.2025.114222}
}

@article{tahir2024hybrid,
  title = {Sustainable hybrid station design framework for electric vehicle charging and hydrogen vehicle refueling based on multiple attributes},
  author = {Tahir, Muhammad Furqan and Hu, Jin and Khan, Abdul and Zhu, Jianguo},
  journal = {Energy Conversion and Management},
  year = {2024},
  doi = {10.1016/j.enconman.2023.117922}
}

@article{liu2024critical,
  title = {Improving critical buildings energy resilience via shared autonomous electric vehicles: A sequential optimization framework},
  author = {Liu, Hongyu and Abdin, Adam and Puchinger, Jakob},
  journal = {Computers \& Operations Research},
  year = {2024},
  doi = {10.1016/j.cor.2023.106513}
}

@article{borrotti2024conformal,
  title = {Quantifying uncertainty with conformal prediction for heating and cooling load forecasting in building performance simulation},
  author = {Borrotti, Matteo},
  journal = {Energies},
  volume = {17},
  number = {17},
  pages = {4348},
  year = {2024},
  doi = {10.3390/en17174348}
}

@article{zhang2023decentralized,
  title = {Novel data-driven decentralized coordination model for electric vehicle aggregator and energy hub entities in multi-energy system using an improved multi-agent DRL approach},
  author = {Zhang, X. and Hu, Z. and Cao, Y. and Ghias, A. M. Y. M.},
  journal = {Applied Energy},
  year = {2023},
  doi = {10.1016/j.apenergy.2023.120902}
}

@article{gokhale2022pinn,
  title = {Physics informed neural networks for control oriented thermal modeling of buildings},
  author = {Gokhale, Gaurav and Claessens, Bert and Develder, Chris},
  journal = {Applied Energy},
  year = {2022},
  doi = {10.1016/j.apenergy.2022.118852}
}

@article{tian2021resilience,
  title = {Energy cost and efficiency analysis of building resilience against power outage by shared parking station for electric vehicles and demand response program},
  author = {Tian, Wei and Talebizadehsardari, Pouya},
  journal = {Energy},
  year = {2021},
  doi = {10.1016/j.energy.2020.119058}
}

@article{lei2018distribution,
  title = {Distribution-free predictive inference for regression},
  author = {Lei, Jing and G'Sell, Max and Rinaldo, Alessandro and Tibshirani, Ryan J. and Wasserman, Larry},
  journal = {Journal of the American Statistical Association},
  volume = {113},
  number = {523},
  pages = {1094--1111},
  year = {2018},
  doi = {10.1080/01621459.2017.1307116}
}

@article{shafer2008tutorial,
  title = {A tutorial on conformal prediction},
  author = {Shafer, Glenn and Vovk, Vladimir},
  journal = {Journal of Machine Learning Research},
  volume = {9},
  pages = {371--421},
  year = {2008},
  url = {https://jmlr.org/papers/v9/shafer08a.html}
}

@inproceedings{gibbs2021adaptive,
  title = {Adaptive conformal inference under distribution shift},
  author = {Gibbs, Isaac and Cand{\`e}s, Emmanuel},
  booktitle = {Advances in Neural Information Processing Systems},
  volume = {34},
  pages = {1660--1672},
  year = {2021},
  url = {https://proceedings.neurips.cc/paper/2021/hash/0d441de75945e5acbc865406fc9a2559-Abstract.html}
}

@article{crawley2001energyplus,
  title = {EnergyPlus: creating a new-generation building energy simulation program},
  author = {Crawley, Drury B. and Lawrie, Linda K. and Winkelmann, Frederick C. and Buhl, W. F. and Huang, Y. Joe and Pedersen, Curtis O. and Strand, Richard K. and Liesen, Richard J. and Fisher, Daniel E. and Witte, Michael J. and Glazer, Jason},
  journal = {Energy and Buildings},
  volume = {33},
  number = {4},
  pages = {319--331},
  year = {2001},
  doi = {10.1016/S0378-7788(00)00114-6}
}

@article{raissi2019pinn,
  title = {Physics-informed neural networks: A deep learning framework for solving forward and inverse problems involving nonlinear partial differential equations},
  author = {Raissi, Maziar and Perdikaris, Paris and Karniadakis, George Em},
  journal = {Journal of Computational Physics},
  volume = {378},
  pages = {686--707},
  year = {2019},
  doi = {10.1016/j.jcp.2018.10.045}
}

@misc{rackauckas2020ude,
  title = {Universal differential equations for scientific machine learning},
  author = {Rackauckas, Christopher and Ma, Yingbo and Martensen, Julius and Warner, Collin and Zubov, Kirill and Supekar, Rohit and Skinner, Dominic and Ramadhan, Ali and Edelman, Alan},
  year = {2020},
  eprint = {2001.04385},
  archivePrefix = {arXiv},
  primaryClass = {cs.LG},
  url = {https://arxiv.org/abs/2001.04385}
}

@misc{energyplus2026,
  title = {EnergyPlus building energy simulation program},
  author = {{U.S. Department of Energy}},
  year = {2026},
  url = {https://energyplus.net/}
}

@misc{actransitdata,
  title = {Data API Resource Center},
  author = {{Alameda-Contra Costa Transit District}},
  year = {2026},
  url = {https://www.actransit.org/data-api-resource-center}
}

@misc{townofcarydata,
  title = {Town of Cary Open Data},
  author = {{Town of Cary}},
  year = {2026},
  url = {https://data.townofcary.org/}
}

@misc{evloadopendata,
  title = {EV Load Open Data},
  author = {Amara, Yvenn},
  year = {2026},
  url = {https://github.com/yvenn-amara/ev-load-open-data}
}

\end{document}